\newcommand{\CLASSINPUTinnersidemargin}{17mm}
\documentclass[10pt,final,journal,letterpaper,oneside,twocolumn]{IEEEtran}
\usepackage[utf8x]{inputenc}
\usepackage{ucs}
\usepackage[english]{babel}
\usepackage[T1]{fontenc}
\usepackage{euscript}
\usepackage{amsmath}
\usepackage{amsfonts}
\usepackage{amssymb}
\usepackage{exscale}
\usepackage{graphics}
\usepackage{graphicx}
\usepackage{battail}
\usepackage{fancyhdr}
\usepackage{mathtext}
\usepackage{latexsym}
\usepackage{fancybox} %: Variants of \fbox
\usepackage{pstricks} %
\usepackage{lineno}
\usepackage{xcolor}
\usepackage{subcaption}
\usepackage{makecell}
\usepackage{wrapfig}

\author{Zijun C. Zhao, Maxim Goryachev, Jerzy Krupka, and Michael E. Tobar
\thanks{Zijun C. Zhao,Maxim Goryachev, Michael E. Tobar are with ARC Centre of Excellence for Engineered Quantum Systems, Department of Physics, University of Western Australia, 35 Stirling Highway, Crawley WA 6009, Australia. Jerzy Krupka is with Instytut Mikroelektroniki i Optoelektroniki PW, Koszykowa 75, 00-662 Warsaw, Poland}}

\title{Precision Multi-Mode Dielectric Characterization of a Crystalline Perovskite Enables Determination of the Temperature-Dependent Phase Transitions}

\begin{document}
\maketitle %\maketitle must follow title, authors, abstract and \pacs

\date{\today}

\begin{abstract}
Simple perovskite crystals undergo structural phase transitions on cooling to low temperatures, which significantly change the material properties of the crystal. In this work we rigorously characterize the temperature evolution of permittivity of a perovskite crystal as it undergoes phase transitions. In particular, we have undertaken precision measurements of a single crystal of Strontium Titanate from 294.6 K to 5.6 K, by measuring the frequency of multiple microwave transverse electric and magnetic resonant modes simultaneously. The multi-mode microwave measurement technique of resonant frequency used in this work allows high precision determination of any induced anisotropy of the permittivity as the crystal undergoes structural phase transitions. Compared with previous results we unequivocally show that the permittivity has an isotropic value of $316.3\pm2.2$ at room temperature, consistent with its well-known cubic structure, and determine the onset of dielectric anisotropy as the crystal is cooled to lower temperatures. We show that the crystal exhibits uniaxial anisotropy in the permittivity below 105 K when the structure becomes tetragonal, and exhibits biaxial anisotropy in the permittivity below 51 K when the structure becomes orthorhombic.
\end{abstract}

\section{Introduction}

Perovskites are inorganic complexes that have a crystalline structure similar to that of the mineral perovskite (calcium titanate,  CaTiO$_3$). Other examples of simple crystalline perovskites include, strontium titanate, lead titanate and barium titanate, which may undergo structural phase transitions as a function of temperature. The behaviour of the emergent macroscopic polarization of such structures is non-trivial, with the possibility of a number of phases, which can depend on a variety of length scales and imperfections such as polar nano clusters, strain history and crystal defects \cite{Lee2005ui,Kalinin2010,Biancoli2015ux,Bencan2021vy,Hofling2021us,Kinbara2007TemperatureDO,7002922,Qiu2020tj}. In general perovskites are found in an enormous number of materials, with a range of applications and properties \cite{ARTINI2017427}. In this work we analyse crystalline Strontium titanate (SrTiO$_3$) or STO, which has been a very popular material for study at low temperatures due to its unique properties, such as extremely large permittivity and the ability to tune the permittivity at low temperatures with a DC bias voltage applied across the crystal. 

Crystalline STO has a cubic perovskite structure at room temperature and as it is cooled, undergoes a well-known cubic to tetragonal structural transition at 105 K. This antiferrodistortive (AFD) phase transition occurs due to the softening of a transverse optical phonon mode, which is very sensitive to axial stress \cite{Muller:1979aa,Cao:2000aa}. Below this temperature the properties of the crystal can depend on how the sample is cooled. For example, prior work shows that if it is stressed or has a DC voltage applied during the cooling process, the properties of the crystal can be significantly affected \cite{Arzel2000,Pesquera2018}. The crystal approaches a ferroelectric phase at 51 K, however quantum fluctuations of the low frequency phonon modes prevent it from completely undergoing the corresponding phase transition, becoming a quantum paraelectric at 5 K \cite{Muller:1979aa}, unless an external stress or electric field is applied \cite{Cao:2000aa,Neville:1972aa,Vendik:1998aa,Hemberger_1996,Sidoruk_2010,Kvasov16, li2019terahertz}.  There have been many studies of the properties of STO down to low temperatures, both theoretically \cite{Yuan:2003aa,Palova:2009aa,Kvasov16,Tagantsev2001} and experimentally \cite{Muller:1979aa, Neville:1972aa,Vendik:1998aa,Muller:1991av, Geyer:2005aa,Trainer:2001aa,Krupka:1994aa,Lytle1964aa,Salje2013,Blinc2005,Arzel_2003}, outlining the complex and interesting variation of permittivity and crystal structure. 

The high permittivity and large dependence of permittivity on applied DC electric field are important characteristics of STO, which allows device miniaturisation and devices that may be tuned as a function of DC bias. Examples include room temperature masers \cite{Breeze:2015aa} and voltage controlled superconducting electronic devices \cite{Lancaster:1998aa}.  Further utilisation of STO in various fields of low temperature physics has lead to extensive studies of its electromagnetic properties. For example, it was recently determined that anisotropy below the AFD phase transition occurs due to a trilinear coupling between the polarisation, AFD lattice tilts and an antiferroelectric (AFE) mode \cite{Casals2018}.  

The measurement of resonance frequency of a dielectric loaded cavity resonators, is a common way to characterise the electromagnetic properties of materials over a broad range of frequencies and temperatures \cite{Le-Floch:2014aa}. More recently, the measurement of multiple modes in a single resonant cavity structure has become a successful technique for the determination of uniaxial \cite{Tobar:1998aa,Krupka:1999aa,Krupka:1999ab} and biaxial permittivity \cite{Carvalho:2015aa,Carvalho2017}, for a range of important single crystals. These techniques can use a combination of low order Transverse Electric (TE) and Transverse Magnetic (TM) modes, as well as higher order modes such as Whispering Gallery Modes (WGM) to successfully characterise the anisotropy. The main advantages of this technique is due to the use of a large number of modes in the centimetre and millimetre wave frequency ranges with varying mode polarisation and high magnetic and electrical filling factors within the sample. Due to these properties WGM devices have also been implemented to characterise material electromagnetic loss and allow the comprehensive Electron Spin Resonance (ESR) spectroscopy of various ions within different crystals using the multiple modes \cite{PhysRevB.88.224426,quartzG,Goryachev:2014ab, Goryachev:2015aa, Creedon:2015aa,Creedon:2011wk,Hosain_2018}. Recently this technique was applied to an STO crystal, which measured extra dielectric losses at microwave frequencies at temperatures below the phase transition temperature at 51 K \cite{Hosain2019}, consistent with previous measurements \cite{Geyer:2005aa, Krupka:1994aa}. 

In this work, we characterise the microwave permittivity of a cylindrical crystalline STO sample, grown along its cylindrical axis, and show that the permittivity is isotropic at room temperature by measuring multiple modes of different electric field polarisations. As the crystal is cooled we unequivocally show the material first becomes uniaxially anisotropic at 105 K and then biaxially anisotropic at 51 K. These results give important information for designing devices over this temperature range, which utilise the unique properties of STO.

\section{Methods}

\subsection{Microwave cavity for STO}
In order to probe the dielectric properties of STO, a cylindrical single crystal specimen of diameter 3.27 mm and height 3.66 mm was utilised. The crystal was put in a cylindrically symmetric oxygen free copper cavity with diameter of 8 mm and height of 4.7 mm. The crystal was placed on top of a sapphire disk substrate and kept in place by a teflon piece on top, as demonstrated in Fig.~\ref{expset} inset.  The c axis of the crystal was aligned with the z axis shown in Fig.~\ref{expset} inset. The sapphire substrate and the teflon piece were utilised in order to uncouple the crystal modes from the cavity walls, and thus decouple the modes from the resistive losses of the copper cavity. Sapphire was chosen on the bottom of the cavity due to its low dielectric loss and good thermal conductivity at low temperatures. On the other hand, these materials do not introduce any significant error to the measurement results due to orders of magnitude difference in permittivity with the STO crystal. Microwave radiation in the cavity was excited via both straight electrode antenna and loop probes separately in order to resolve the mode polarisations (i.e. TE or TM). 

\subsection{Finite element simulation} \label{SectionSimulation}

\begin{figure}[ht!]
	\centering
			\includegraphics[width=0.46\textwidth]{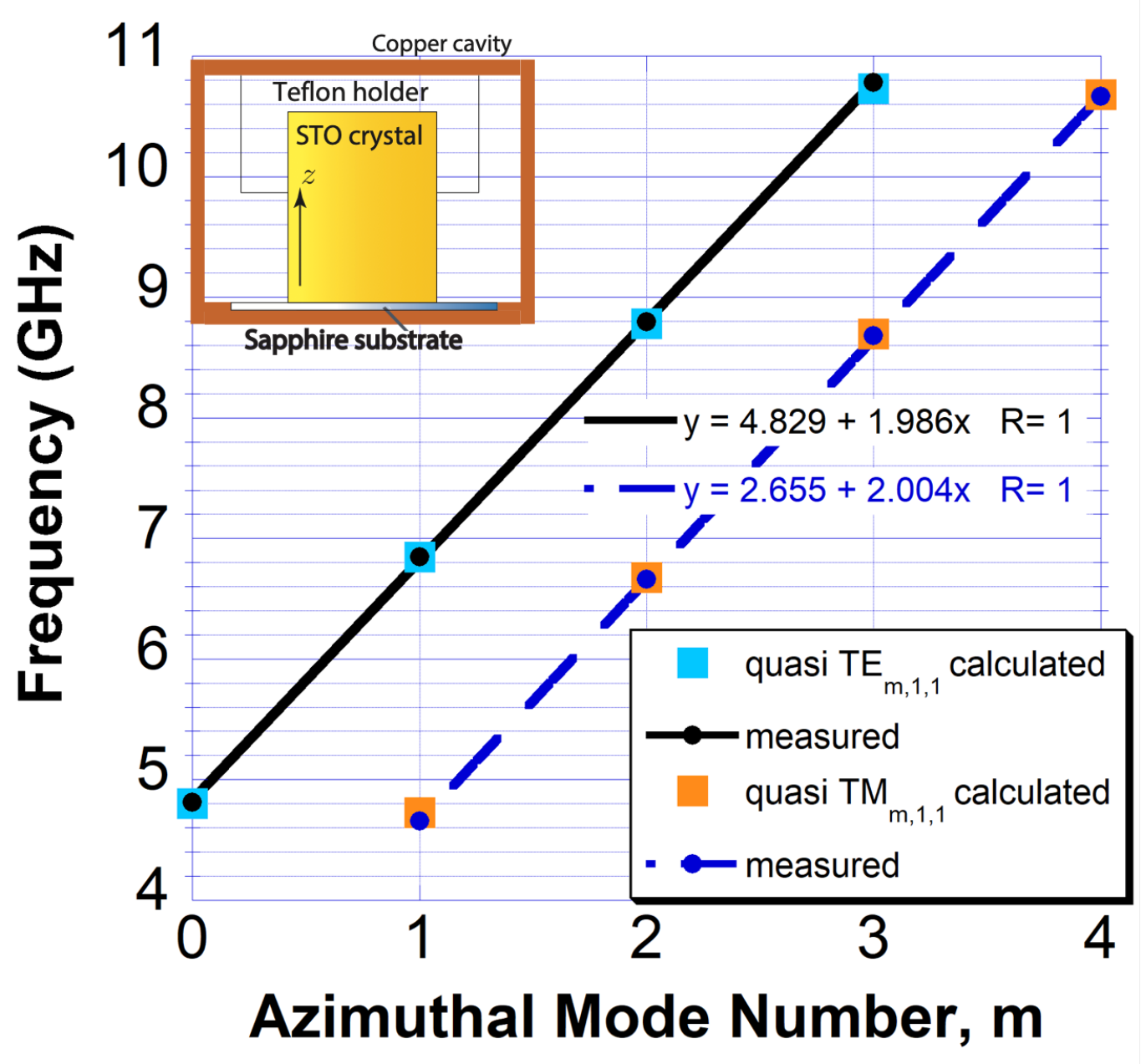}
	\caption{Comparison between measured and calculated frequencies of the fundamental quasi-TE and TM mode families with the value of permittivity of STO calculated to be $316.3$. Inset: Side view or $(r,z)$ plane of the STO cylindrical cavity. The cylindrical crystal is mounted on a sapphire disc substrate and held in place by a teflon piece.}
	\label{expset}
\end{figure}

To calculate the resonant mode frequencies for comparison with experimental measurements, finite element simulation by COMSOL Multiphysics was implemented.  Because of the cylindrical symmetry of the cavity a two dimensional mesh was prepared representing the cavity as shown in Fig.~\ref{expset}. To avoid significant mesh errors, the inbuilt ultra-fine meshing function was selected. The model utilised known dielectric parameters of sapphire and teflon and assumed perfect conducting walls for the cavity. To calculate material properties of STO across the broad temperature range, the permittivity was iterated in the model, to minimise the error between the calculated and measured frequencies for the selected modes. 

In general, the permittivity of STO can be represented by a tensor of the following biaxial form:
\begin{equation}
\begin{pmatrix}
\varepsilon_x & 0 & 0 \\
0 & \varepsilon_y & 0 \\
0 & 0 & \varepsilon_z \\
\end{pmatrix}
\label{tensor}
\end{equation}
where the permittivity in general is different along the $x$, $y$ and $z$ coordinates in the crystal, with the cylindrical $z$-axis aligned with the crystal $c$-axis. The electromagnetic modes couple to the permittivity tensor, depending on the electrical energy filling factors, $Pe_i $, in the $i=x$, $y$ and $z$ directions, which are given by \cite{Krupka:1999ab},
\begin{equation}
Pe_i = \frac{\iiint_{V_{d}} {\varepsilon_{i}\mathbf{E}_i \cdot \mathbf{E}_i^{*} dv}}{\iiint_{V} {\varepsilon(v)\mathbf{E} \cdot \mathbf{E^{*}} dv}}.
\label{FF}
\end{equation}
Here, $\mathbf{E}_i$ is the vector component of the electric field in the $i$ direction within the sample volume, $V_d$, which in this case is the STO crystal, and $\mathbf{E}=E_x\hat{x}+E_y\hat{y}+E_x\hat{z}$, which is the total electric field integrated over the entire volume, $V$. If a material is biaxial anisotropic in permittivity, at least three distinct modes are needed to resolve the different values of $\varepsilon_{i}$. 

Because the STO crystal is isotropic at room temperature and the permittivity increases as the cavity is cooled, our process calculated the permittivity and filling factors at room temperature first. Following this we obtained a rough estimate for the increasing value of permittivity with temperature based on the measured negative frequency shift of all modes and the predicted change in all mode filling factors from equations derived in \cite{Krupka:1999ab}, which relate the temperature dependence of mode frequency to the temperature dependence of material permittivity and the calculated electric energy filling factors. From this initial estimate, the permittivity tensor was iterated in a rigorous way to obtain the best results by minimising the differences between the simulated and experimental mode frequencies. 

\begin{figure*}[ht!]
	\centering
			\includegraphics[width=0.94\textwidth]{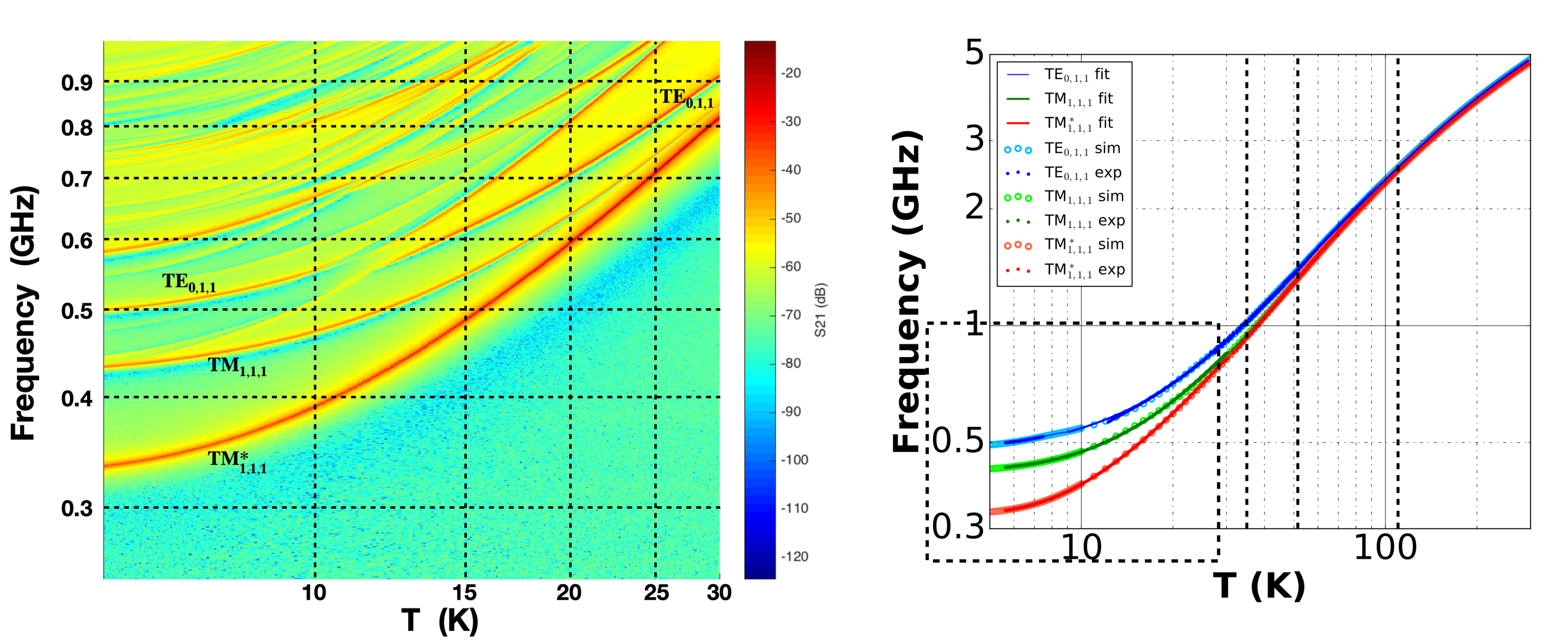}
	\caption{Left: Color density plot of experimental measurements of $S_{21}$ versus temperature below 30 K, wth the TE$_{0,1,1}$ mode and the quasi TM$_{1,1,1}$ / TM$^*_{1,1,1}$ modes identified. The discontinuity of TE$_{0,1,1}$ around 12 K is due to the mode crossing caused by anisotropy at low temperatures of STO, which leads to different temperature coefficients for the TM and TE modes. Right: Frequency versus temperature for the TE$_{0,1,1}$ mode and the quasi TM$_{1,1,1}$ / TM$^*_{1,1,1}$ modes. The solid dots represent experimental measurements of frequency, while the line represents polynomial fits. The hollow circles represent results from the finite element simulation. Due to the high permittivity of the STO, electrical energy is mostly confined within the crystal with filling factors of each mode summing up close to unity (or 100\%) across the whole temperature range.}
	\label{freqvsT}
\end{figure*}

\subsection{Precision frequency measurements at room and cryogenic temperatures}

The system was characterised in transmission ($S_{21}$) with a vector network analyser (VNA) referenced to a high stability atomic hydrogen maser. At room temperature, resonance mode frequencies were found by searching manually in the range of 4 to 20 GHz and identified by comparison with the simulation, which are shown in Fig.~\ref{expset}. For the considered structure, only modes with $m=0$ can be pure TM or TE polarisation, for $m>1$ the modes are quasi TE and TM and become WGM as $m$ increases. The lowest four modes in frequency of the fundamental quasi TE and TM mode families were measured with azimuthal mode numbers of up to $m = 4$. 

For cryogenic measurements, the copper cavity was attached to the 4 K plate of a cryogen free pulsed tube system and measured in $S_{21}$ with VNA. In order to measure the temperature dependence, a Cernox calibrated thermometer was  attached to the copper cavity and continuously measured throughout the experiment. Measurements were implemented by warming up and cooling down very slowly in thermal equilibrium, with temperature lag between the thermometer and crystal kept to a minimum. The lowest frequency TE and TM modes were identified from the temperature measurements and used for the permittivity calculations. 

\section{Results and discussion}

\subsection{Characterisation at Room Temperature}

At room temperature, it is well known that STO exhibits a cubic structure and therefore is most likely isotropic. Despite this, early work by Neville et al. that implemented a capacitive technique using single crystal wafers between 0.125 mm and 1 mm thickness in various crystal orientations, reported anisotropy at room temperature \cite{Neville:1972aa}. In their experiment, gold electrodes were deposited on the samples and the permittivity was inferred from 4.2 K to 300 K and between 1 kHz to 50 MHz. Their permittivity measurements yield 330 in the [001] direction, 458 in the [011] direction and 448 in the [111] direction at room temperature. This type of scatter in measurements with only a $\pm$ 2\% error and is not consistent with our isotropic value of permittivity presented in this work. In the next section we compare our results across the whole temperature range and discuss the most likely discrepancies between our measurements and the ones presented in the literature\cite{Neville:1972aa}.

\subsection{Revealing Crystal Anisotropy at Cryogenic Temperatures}

\begin{figure*}[ht!]
	\centering
			\includegraphics[width=0.94\textwidth]{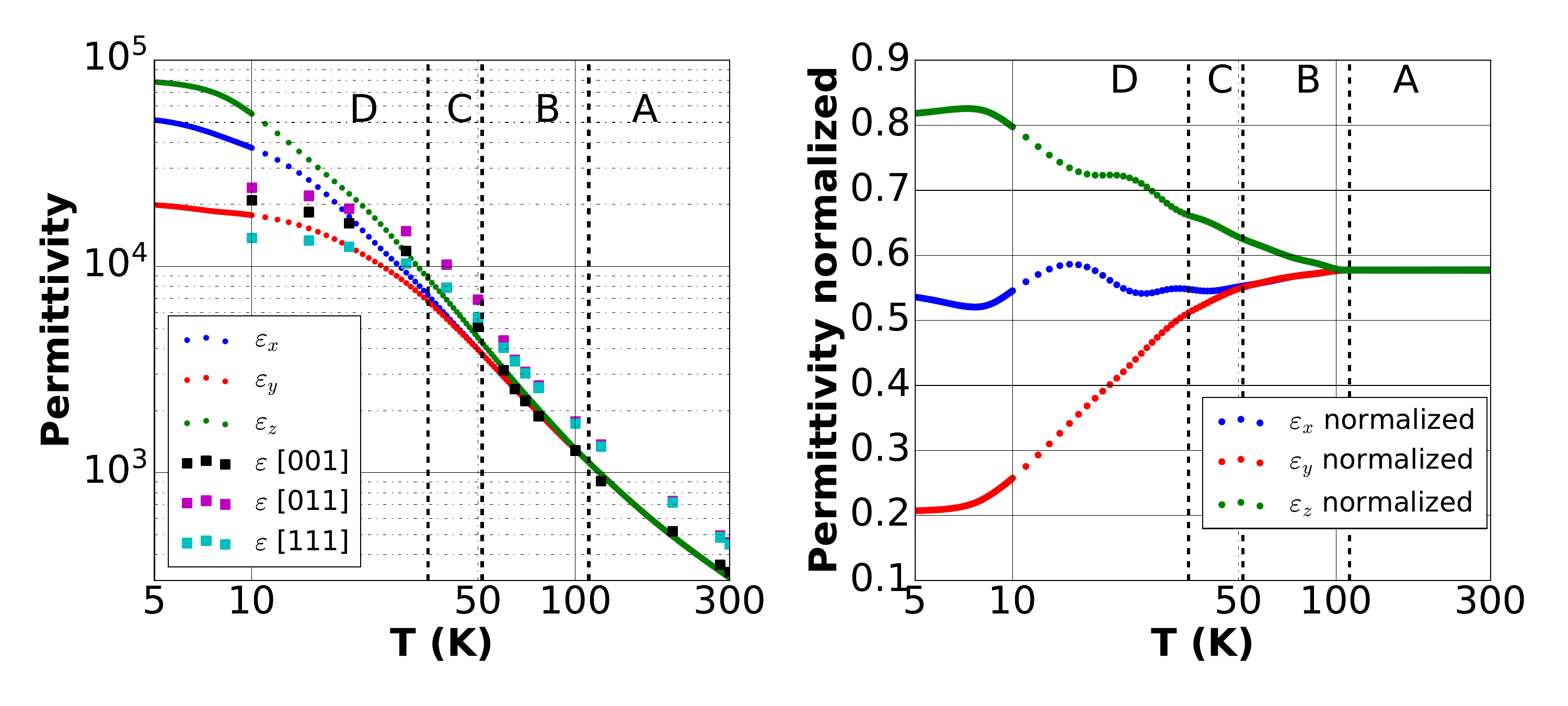}
	\caption{Left: Derived permittivities $\varepsilon_{x}$, $\varepsilon_{y}$ and $\varepsilon_{z}$ from the frequency measurement in Fig.~\ref{freqvsT} and the finite element simulation, with comparison to measurements from the literature\cite{Neville:1972aa} along different crystallographic directions. Right: Normalized permittivities $\varepsilon_{x}$, $\varepsilon_{y}$ and $\varepsilon_{z}$ versus temperature. Each permittivity is normalized by dividing a value of $\sqrt{\varepsilon_{x}^{2}+{\varepsilon_{y}^{2}}+\varepsilon_{z}^{2}}$.}
	\label{anisotropy}
\end{figure*}

Fig.~\ref{expset} shows clearly good agreement for the fundamental TM and TE mode families, with a calculated isotropic permittivity of $316.3$, which minimises the aggregate error between simulation and experiment. Calculations gave an average error from the four TE modes of $0.18\%$, while the average of the four TM modes was determined to be $0.52\%$. Also, dimensions were measured to the nearest hundredth of a mm using laser technique giving approximately $0.3\%$ error in height and radius. Adding all errors in quadrature gives a value of permittivity of $316.3\pm2.2$, consistent with more recent measurements of permittivity, which only implemented TE modes \cite{Krupka:1994aa,Breeze:2015aa}. The mode free-spectral range was measured to be nearly constant and close to 2 GHz and the permittivity was demonstrated to be effectively constant with no frequency dependence observed between 4 to 11 GHz. This result is consistent with the theory of the soft mode phonon, which dominates the value of permittivity in this frequency range \cite{Neville:1972aa}. Prior work predicts the permittivity to be constant beyond a THz at room temperature and beyond 100 GHz at 4.2 K \cite{Neville:1972aa}.

The frequency versus temperature measurements are shown in Fig.~\ref{freqvsT}. The lowest frequency modes were measured and identified as the TE$_{0,1,1}$ mode and the quasi TM$_{1,1,1}$ / TM$^*_{1,1,1}$ doublet mode pair. As the temperature was reduced, the TM and TE modes diverge in frequency below 110 K, indicating uniaxial anisotropy, and then below 51 K the TM$_{1,1,1}$ mode doublet degeneracy is lifted. Recently it was shown that the doublet degeneracy will be lifted if the permittivity is biaxial, and this effect was used to characterise biaxial material \cite{Carvalho:2015aa}. It is clear from the temperature dependence of the observed mode frequencies as shown in Fig.~\ref{freqvsT},  that the material exhibits uniaxial anisotropy below 105 K, and then biaxial anisotropy in permittivity below 51 K.

Due to the high dielectric constant of the STO, the electrical energy is nearly $100\%$ confined within the crystal and thus the sum of the electric energy filling factors components of the whole crystal are approximately constant across a wide temperature range and very close to unity. However, as the temperature changes, the individual components of the filling factors also changed. Filling factors were calculated from the COMSOL simulation for each mode. At room temperature, for the TM$_{1,1,1}$ mode the electrical energy filling factor parallel to the c-axis was calculated to be, $Pe_{||}\approx 63\%$ and perpendicular is $Pe_{\bot}\approx37\%$, while for the TE$_{0,1,1}$ mode they were calculated to be, $Pe_{||}= 0\%$ and $Pe_{\bot}\approx100\%$.   At 5.6 K, for the filling factors for the TM$_{1,1,1}$ mode become, $Pe_{||}\approx 35\%$ and $Pe_{\bot}\approx 65\%$ and for the TM$^*_{1,1,1}$ mode they become, $Pe_{||}\approx 51\%$ and $Pe_{\bot}\approx 49\%$, while for the TE$_{0,1,1}$ mode they became, $Pe_{||}\approx 4\%$ and $Pe_{\bot}\approx 96\%$, so in some cases was no longer a pure TE mode due to the hybridization with a higher order hybrid mode when a frequency mode crossing occurred. Previous measurements that utilised TE modes of $m=0$ only determined the perpendicular component of permittivity \cite{Geyer:2005aa,Krupka:1994aa}. In this work, the anisotropy was revealed due to the simultaneous measurement of multiple modes as shown in Fig. {\ref{freqvsT}. Thus, through numerical iteration all the tensor components of permittivity were found as explained in Section \ref{SectionSimulation}. With the additional data of TM$^*_{1,1,1}$ mode, the anisotropy in perpendicular component of the permittivity was also revealed. 

The values of permittivity derived from the curve fits in Fig.~\ref{freqvsT} and the finite element modelling are shown in Fig.~\ref{anisotropy} were compared with data from the literature \cite{Neville:1972aa}. Our results show a significant change in anisotropy, which can be explained by crystal structure phase transitions. Plots in Fig.~\ref{anisotropy} are subdivided into sections A-D in accordance with corresponding phases. From room temperature down to 110 K, STO exhibits highly symmetric cubic perovskite structure (space group \textit{Pm3m}) as demonstrated in section A. For lower temperatures, the crystal structure is deformed from cubic by a number of effects. STO demonstrates tetragonal (space group \textit{I4/mcm}) structure in the range $110-51$ K (section B) shifting to orthorhombic in the range $51-35$ K (section C).  It has also been demonstrated that STO could form a single low-symmetry phase below 35 K (section D), which is possibly a rhombohedral structure \cite{Lytle1964aa}. 

These temperature-dependent cubic-tetragonal-orthohomic structural changes of STO have been observed experimentally by X-Ray diffraction \cite{Lytle1964aa}. Similar cubic-tetragonal-orthohomic phase transitions of STO through crystal structural changes under high pressures up to 26 GPa have also been observed using Raman Spectroscopy\cite{grzechnik1997raman} and confirmed using \textit{ab initio} density-functional-theory calculations\cite{hachemi2010elasticity}. It is evident from Fig.~\ref{anisotropy} that the main transition to biaxial anisotropy occurs when the crystal structure shifts to the orthorhombic phase near 50 K in temperature. This coincides with a maxima of temperature-dependent quality factor of the quasi-TE$_{2,1,1}$ TM$_{2,1,1}$ as reported in \cite{Hosain2019}. There is evidence in the literature that suggests that the structural changes of STO are not due to the bulk properties of the material, but due to the increasing polarity of ferroelastic twin walls, and ferroelastic domain walls in STO become ferroelectric, which leads to the phase transtion near 50 K \cite{Pesquera2018,Salje2013, scott2012domain}. In this case, the local ferroelectricity are domains atomistically thin based on the size of the twin walls. However, it is not clear how such local ferroelectricity caused by domain walls can explain the macroscopic anisotropy observed here in the STO crystal. 

The results we measured do not show any clear anisotropy at room temperature as originally claimed in the literature \cite{Neville:1972aa}.  The varying results are most likely due to systematic scatter of data caused by dimensional and other errors in their measurements. Also, below 50 K the previous results \cite{Neville:1972aa} are not clearly anisotropic either, and in general give a lower value of permittivity below 15 K than those measured in this work. It has been shown that the application of stress to the crystal lowers the permittivity of the crystal \cite{Muller:1979aa}, and that the permittivity of STO is highly stress dependent. We postulate that the results in the literature \cite{Neville:1972aa} had excess stress created in the crystal due to the differential contraction of the gold electrodes deposited on the crystal. Our technique in contrast is stress free as the crystal sits on a sapphire substrate, with an indentation slightly bigger than the crystal radius to keep the crystal in place, along with a loosely fitting teflon cap held from the top of the cavity. 

\section{Conclusions}
In summary, precision measurement of the permittivity was undertaken by continuously measuring multiple TE and TM resonant modes within a single crystal dielectrically loaded microwave cavity resonator in thermal equilibrium from 5.6 K to room temperature. It was shown that the permittivity is isotropic at room temperature with a value of $316.3\pm2.2$ by measuring multiple modes of different electric field polarisations. The SrTiO$_3$ crystal underwent two displacive phase transitions around 105 K and 51 K. The temperature dependent permittivity obtained for the SrTiO$_3$ crystal was consistent with the known transformations of the crystal structure during these transitions. The crystal became uniaxially anisotropic below the antiferrodistortive phase transition at 105 K, when the lattice transforms from cubic to tetragonal, and then biaxially anisotropic below the phase transition at 51 K, when the lattice transforms from tetragonal to orthorhombic. We have shown that precision measurements of the frequency of multiple modes is a practical technique to study phase transitions in high permittivity ferroelectric crystals.

\section*{Acknowledgements}
We thank Prof. Andrew Johnson for useful discussions regarding this work, we also thank Dr. Jeremy Bourhill and graduate students Lingfei Zhao and Boyang Jiang for assistance with some of the data analysis. This work was supported by the Australian Research Council Grant No. CE170100009.

\ifCLASSOPTIONcaptionsoff
  \newpage
\fi

\end{document}